\documentstyle[prb,aps,twocolumn,graphicx]{revtex}

\newcommand{\ket}[1]{|#1\rangle}

\newcommand{\sfrac}[2]{
    \textstyle
    \frac{#1}{#2}
    \displaystyle}

\begin{document}

\title{Quantum Cryptography with Entangled Photons}

    \author{Thomas Jennewein, Christoph Simon, Gregor Weihs, \\
    Harald Weinfurter\dag, and Anton Zeilinger}

    \address{Institut f\"{u}r Experimentalphysik, Universit\"{a}t Wien, \\
    Boltzmanngasse 5, A--1090  Wien, Austria \\
   \dag Sektion Physik, Universit\"{a}t M\"{u}nchen, \\
    Schellingstr. 4/III, D-80799 M\"{u}nchen, Germany\ddag }

\date{\today}

\maketitle

\thispagestyle{empty}

\begin{abstract}

By realizing a quantum cryptography system based on polarization
entangled photon pairs we establish highly secure keys, because a
single photon source is approximated and the inherent randomness
of quantum measurements is exploited. We implement a novel key
distribution scheme using Wigner's inequality to test the security
of the quantum channel, and, alternatively, realize a variant of
the BB84 protocol. Our system has two completely independent users
separated by $360$~m, and generates raw keys at rates of $400$ --
$800$~bits/second with bit error rates arround $3$\%.

\end{abstract}



\vspace*{1cm}

The primary task of cryptography is to enable two parties
(commonly called Alice and Bob) to mask confidential messages
such, that the transmitted data are illegible to any unauthorized
third party (called Eve). Usually this is done using shared secret
keys. However, in principle it is always possible to intercept
classical key distribution unnoticedly. The recent development of
quantum key distribution \cite{BB84} can cover this major loophole
of classical cryptography. It allows Alice and Bob to establish
two completely secure keys by transmitting single quanta (qubits)
along a quantum channel. The underlying principle of quantum key
distribution is that nature prohibits to gain information on the
state of a quantum system without disturbing it. Therefore, in
appropriately designed schemes, no tapping of the qubits is
possible without showing up to Alice and Bob. These secure keys
can be used in a One-Time-Pad protocol \cite{Vernam}, which makes
the entire communication absolutely secure.

Two well known concepts for quantum key distribution are the BB84
scheme and the Ekert scheme. The BB84 scheme \cite{BB84} uses
single photons transmitted from Alice to Bob, which are prepared
at random in four partly orthogonal polarization states:
$0^\circ$, $45^\circ$, $90^\circ$, $135^\circ$. If Eve tries to
extract information about the polarization of the photons she will
inevitably introduce errors, which Alice and Bob can detect by
comparing a random subset of the generated keys.

The Ekert scheme \cite{Ekert} is based on entangled pairs and uses
Bell's inequality \cite{Bell} to establish security. Both Alice
and Bob receive one particle out of an entangled pair. They
perform measurements along at least three different directions on
each side, where measurements along parallel axes are used for key
generation and oblique angles are used for testing the inequality.
In \cite{Ekert}, Ekert pointed out that eavesdropping inevitably
affects the entanglement between the two constituents of a pair
and therefore reduces the degree of violation of Bell's
inequality. While we are not aware of a general proof that the
violation of a Bell inequality implies the security of the system,
this has been shown \cite{otherprot} for the BB84 protocol adapted
to entangled pairs and the CHSH inequality \cite{chsh}.

In any real cryptography system, the raw key generated by Alice
and Bob contains errors, which have to be corrected by classical
error correction \cite{errorcorr} over a public channel.
Furthermore it has been shown that whenever Alice and Bob share a
sufficiently secure key, they can enhance its security by privacy
amplification techniques \cite{privacy}, which allow them to
distill a key of a desired security level.

A range of experiments have demonstrated the feasibility of
quantum key distribution, including realizations using the
polarization of photons \cite{polexp} or the phase of photons in
long interferometers \cite{phaseexp}. These experiments have a
common problem: the sources of the photons are attenuated laser
pulses which have a non-vanishing probability to contain two or
more photons, leaving such systems prone to the so called beam
splitter attack \cite{cohpulse}.

Using photon pairs as produced by parametric down-conversion
allows us to approximate a conditional single photon source
\cite{singlephot} with a very low probability for generating two
pairs simultaneously and a high bit rate \cite{twopairs}.
Moreover, when utilizing entangled photon pairs one immediately
profits from the inherent randomness of quantum mechanical
observations leading to purely random keys.

Various experiments with entangled photon pairs have already
demonstrated that entanglement can be preserved over distances as
large as 10~km \cite{entexp}, yet none of these experiments was a
full quantum cryptography system. We present in this paper a
complete implementation of quantum cryptography with two users,
separated and independent of each other in terms of Einstein
locality and exploiting the features of entangled photon pairs for
generating highly secure keys.

In the following we will describe the variants of the Ekert scheme
and of the BB84 scheme which we both implemented in our
experiment, based on polarization entangled photon pairs in the
singlet state
\begin{equation}
\ket{\Psi^-}=\frac{1}{\sqrt{2}}[\ket{H}_A\ket{V}_B-\ket{V}_A\ket{H}_B]
\:,
\end{equation}
where photon $A$ is sent to Alice and photon $B$ is sent to Bob,
and $H$ and $V$ denote the horizontal and vertical linear
polarization respectively. This state shows perfect
anticorrelation for polarization measurements along parallel but
arbitrary axes. However, the actual outcome of an individual
measurement on each photon is inherently random. These perfect
anticorrelations can be used for generating the keys, yet the
security of the quantum channel remains to be ascertained by
implementing a suitable procedure.

Our first scheme utilizes Wigner's inequality \cite{Wigner} for
establishing the security of the quantum channel, in analogy to
the Ekert scheme which uses the CHSH inequality. Here Alice
chooses between two polarization measurements along the axes
$\chi$ and $\psi$, with the possible results $+1$ and $-1$, on
photon $A$ and Bob between measurements along $\psi$ and $\omega$
on photon $B$. Polarization parallel to the analyzer axis
corresponds to a $+1$ result, and polarization orthogonal to the
analyzer axis corresponds to $-1$.

Assuming that the photons carry preassigned values determining the
outcomes of the measurements $\chi, \psi, \omega$ and also
assuming perfect anticorrelations for measurements along parallel
axes, it follows, that the probabilities for obtaining $+1$ on
both sides, $p_{++}$, must obey Wigner's inequality:
\begin{equation}
p_{++}(\chi,\psi) + p_{++}(\psi,\omega) - p_{++}(\chi,\omega) \geq
0 \:. \label{w_ineq}
\end{equation}

The quantum mechanical prediction $p^{qm}_{++}$ for these
probabilities at arbitrary analyzer settings $\alpha$ (Alice) and
$\beta$ (Bob) measuring the $\Psi^-$ state is
\begin{equation}
p^{qm}_{++}(\alpha,\beta)=\sfrac{1}{2} \sin^2 \left( \alpha-\beta
\right) \:.
\end{equation}

The analyzer settings $\chi=-30^\circ$, $\psi=0^\circ$, and
$\omega=30^\circ$ lead to a maximum violation of Wigner's
inequality~(\ref{w_ineq}):
\begin{eqnarray}
& & p^{qm}_{++}(-30^\circ,0^\circ) + p^{qm}_{++}(0^\circ,30^\circ)
- p^{qm}_{++}(-30^\circ,30^\circ) = \nonumber \\ & & =
\sfrac{1}{8}+\sfrac{1}{8}-\sfrac{3}{8} = -\sfrac{1}{8} \geq 0 \:.
\end{eqnarray}

As Wigner's inequality is derived assuming perfect
anticorrelations, which are only approximately realized in any
practical situation, one should be cautious in applying it to test
the security of a cryptography scheme. When the deviation from
perfect anticorrelations is substantial, Wigner's inequality has
to be replaced by an adapted version \cite{ryff}.

\begin{figure}

\includegraphics[width=\columnwidth]{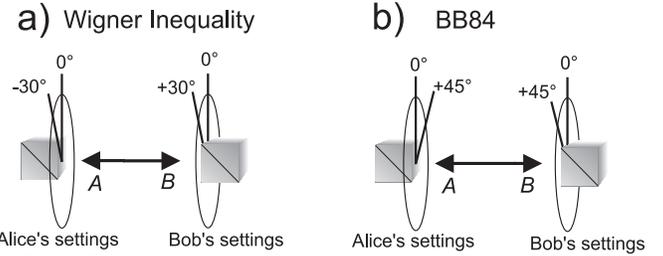}

\caption{Settings for Alice's and Bob's analyzers for realizing
quantum key distribution based either on (a) Wigner's inequality
or (b) the BB84 protocol. The angular coordinates are referenced
to the propagation direction of the particle.}

\label{settings}

\end{figure}

In order to implement quantum key distribution, Alice and Bob each
vary their analyzers randomly between two settings, Alice:
$-30^\circ,0^\circ$  and Bob: $0^\circ,30^\circ$
(Figure~\ref{settings}a). Because Alice and Bob operate
independently, four possible combinations of analyzer settings
will occur, of which the three oblique settings allow a test of
Wigner's inequality and the remaining combination of parallel
settings (Alice$=0^\circ$ and Bob$=0^\circ$) allows the generation
of keys via the perfect anticorrelations, where either Alice or
Bob has to invert all bits of the key to obtain identical keys.

If the measured probabilities violate Wigner's inequality, then
the security of the quantum channel is ascertained, and the
generated keys can readily be used. This scheme is an improvement
on the Ekert scheme which uses the CHSH inequality and requires
three settings of Alice's and Bob's analyzers for testing the
inequality and generating the keys. From the resulting nine
combinations of settings, four are taken for testing the
inequality, two are used for building the keys and three are
omitted at all. However in our scheme each user only needs two
analyzer settings and the detected photons are used more
efficiently, thus allowing a significantly simplified experimental
implementation of the quantum key distribution.

As a second quantum cryptography scheme we implemented a variant
of the BB84 protocol with entangled photons, as proposed in
Reference \cite{BBM}. In this case, Alice and Bob randomly vary
their analysis directions between $0^\circ$ and $45^\circ$
(Figure~\ref{settings}b). Alice and Bob observe perfect
anticorrelations of their measurements whenever they happen to
have parallel oriented polarizers, leading to bitwise
complementary keys. Alice and Bob obtain identical keys if one of
them inverts all bits of the key. Polarization entangled photon
pairs offer a means to approximate a single photon situation.
Whenever Alice makes a measurement on photon $A$, photon $B$ is
projected into the orthogonal state which is then analyzed by Bob,
or vice versa. After collecting the keys, Alice and Bob
authenticate their keys by openly comparing a small subset of
their keys and evaluating the bit error rate.

\begin{figure}

\includegraphics[width=\columnwidth]{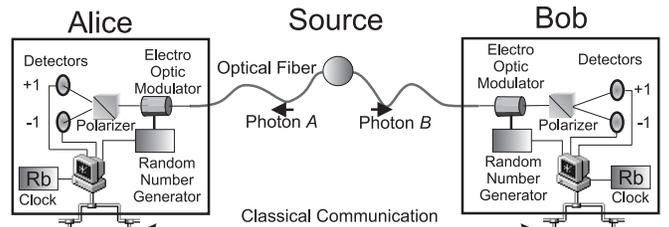}

\caption{The polarization entangled photons are transmitted via
optical fibers to Alice and Bob, who are separated by $360$~m, and
both photons are analyzed, detected and registered independently.
After a measurement run the keys are established by Alice and Bob
through classical communication over a standard computer network.}

\label{setup}

\end{figure}

The experimental realization of our quantum key distribution
system is sketched in Figure~\ref{setup}. Type-II parametric
down-conversion in $\beta$-barium borate \cite{kwiat} (BBO),
pumped with an argon-ion laser working at a wavelength of $351$~nm
and a power of $350$~mW, leads to the production of polarization
entangled photon pairs at a wavelength of $702$~nm. The photons
are each coupled into $500$~m long optical fibers and transmitted
to Alice and Bob respectively, who are separated by $360$~m.

Alice and Bob both have Wollaston polarizing beam splitters as
polarization analyzers. We will associate a detection of parallel
polarization ($+1$) with the key bit 1 and orthogonal detection
($-1$) with the key bit 0. Electro-optic modulators in front of
the analyzers rapidly switch (rise time $<15$~ns, minimum
switching interval $100$ ns) the axis of the analyzer between two
desired orientations, controlled by quantum random signal
generators \cite{rngpaper}. These quantum random signal generators
are based on the quantum mechanical process of splitting a beam of
photons and have a correlation time of less than $100$~ns.

The photons are detected in silicon avalanche photo diodes
\cite{cova}. Time interval analyzers on local personal computers
register all detection events as time stamps together with the
setting of the analyzers and the detection result. A measurement
run is initiated by a pulse from a separate laser diode sent from
the source to Alice and Bob via a second optical fiber. Only after
a measurement run is completed, Alice and Bob compare their lists
of detections to extract the coincidences. In order to record the
detection events very accurately, the time bases in Alice's and
Bob's time interval analyzers are controlled by two rubidium
oscillators. The stability of each time base is better than 1~ns
for one minute. The maximal duration of a measurement is limited
by the amount of memory in the personal computers (typically one
minute).

Overall our system has a measured total coincidence rate of $\sim
1700 \mathrm{s}^{-1}$ , and a singles rate of $\sim 35 000
\mathrm{s}^{-1}$ . From this, one can estimate the overall
detection efficiency of each photon path to be 5~\% and the pair
production rate to be $7\cdot 10^5 \mathrm{s}^{-1}$. Our system is
very immune against a beam splitter attack because the ratio of
two-pair events is only $\sim 3 \cdot 10^{-3}$, where a two-pair
event is the emission of two pairs within the coincidence window
of $4$~ns. The coincidence window in our experiment is limited by
the time resolution of our detectors and electronics, but in
principle it could be reduced to the coherence time of the
photons, which is usually of the order of picoseconds.

In realizing the quantum key distribution based on Wigner's
inequality, Alice's analyzer switch randomly with equal frequency
between $-30^{\circ}$ and $0^{\circ}$, and Bob's analyzer between
$0^{\circ}$ and $30^{\circ}$. After a measurement, Alice and Bob
extract the coincidences for the combinations of settings of
$(-30^{\circ},30^{\circ})$, $(-30^{\circ},0^{\circ})$ and
$(0^{\circ},30^{\circ})$, and calculate each probability. E.g. the
probability $p_{++}(0^\circ,30^\circ)$ is calculated from the
numbers of coincident events $C_{++}$, $C_{+-}$, $C_{-+}$,
$C_{--}$ measured for this combination of settings by
\begin{equation}
p_{++}(0^\circ,30^\circ)= \frac{C_{++}} {C_{++}+C_{+-}+C_{-+}+
C_{--}}.
\end{equation}
We observed in our experiment that the left hand side of
inequality (\ref{w_ineq}) evaluated to $-0.112 \pm 0.014$. This
violation of (\ref{w_ineq}) is in good agreement with the
prediction of quantum mechanics and ensures the security of the
key distribution. Hence the coincident detections obtained at the
parallel settings $(0^{\circ},0^{\circ})$, which occur in a
quarter of all events, can be used as keys. In the experiment
Alice and Bob established $2162$~bits raw keys at a rate of
$420$~bits/second \cite{bias}, and observed a quantum bit error
rate of $3.4$~\%.

In our realization of the BB84 scheme, Alice's and Bob's analyzers
both switch randomly between $0^{\circ}$ and $45^{\circ}$. After a
measurement run, Alice and Bob extract the coincidences measured
with parallel analyzers, $(0^{\circ},0^{\circ})$ and
$(45^{\circ},45^{\circ})$, which occur in half of the cases, and
generate the raw keys. Alice and Bob collected $\sim 80000$ bits
of key at a rate of $850$ bits/second, and observed a quantum bit
error rate of $2.5$~\%, which ensures the security of the quantum
channel.

For correcting the remaining errors while maintaining the secrecy
of the key, various classical error correction and privacy
amplification schemes have been developed \cite{errorcorr}. We
implemented a simple error reduction scheme requiring only little
communication between Alice and Bob. Alice and Bob arrange their
keys in blocks of $n$ bits and evaluate the bit parity of the
blocks (a single bit indicating an odd or even number of ones in
the block). The parities are compared in public, and the blocks
with agreeing parities are kept after discarding one bit per block
\cite{discard}. Since parity checks only reveal odd occurrences of
bit errors, a fraction of errors remains. The optimal block length
$n$ is determined by a compromise between key losses and remaining
bit errors. For a bit error rate $p$ the probability for $k$ wrong
bits in a block of $n$ bits is given by the binomial distribution
$P_n(k)= {n \choose k} p^k (1-p)^{n-k}$.

\begin{figure}
\includegraphics[width=\columnwidth]{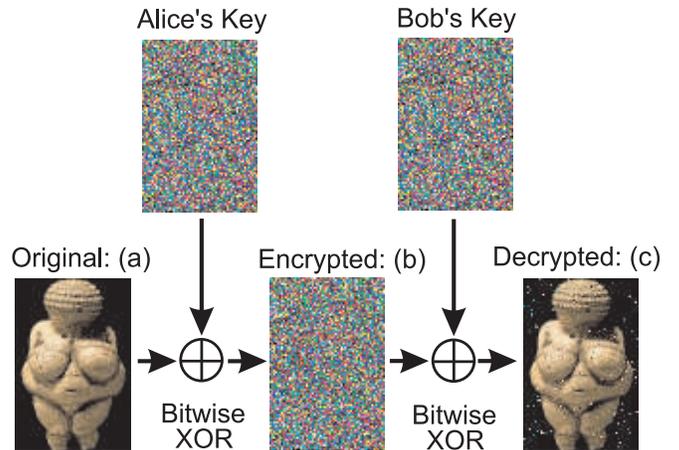}

 \caption{The $49984$ bit large keys generated by the BB84 scheme are used to
 securely transmit an image \protect\cite{Bitmap} (a) of the
``Venus von Willendorf''\protect\cite{Venusdate} effigy. Alice
encrypts the image via bitwise XOR operation with her key and
transmits the encrypted image (b) to Bob via the computer
network. Bob decrypts the image with his key, resulting in (c)
which shows only few errors due to the remaining bit errors in
the keys.}

\label{venus}

\end{figure}

Neglecting terms for three or more errors and accounting for the
loss of one bit per agreeing parity, this algorithm has an
efficiency
$\eta(n)=(1-P_n(1))(n-1)/n$,
defined as the ratio between the key sizes after parity check and
before. Finally, under the same approximation as above, the
remaining bit error rate $p'$ is
$p'=  (1-P_n(0)-P_n(1))(2/n)$.
Our key has a bit error rate $p=2.5$~\%, for which $\eta(n)$ is
maximized at $n=8$ with $\eta(8)=0.7284$, resulting in
$p'=0.40$~\%. Hence, from $\sim 80000$~bits of raw key with a
quantum bit error rate of $2.5$~\%, Alice and Bob use $10$~\% of
the key for checking the security and the remaining $90$~\% of the
key to distill $49984$ bits of error corrected key with a bit
error rate of $0.4$\%. Finally, Alice transmits a $43200$ bit
large image to Bob via the One-Time-Pad protocol, utilizing a
bitwise XOR combination of message and key data
(Figure~\ref{venus}).

In this letter we presented the first full implementation of
entangled state quantum cryptography. All the equipment of the
source and of Alice and Bob has proven to operate outside shielded
lab-environments with a very high reliability. While further
practical and theoretical investigations are still necessary, we
believe that this work demonstrates that entanglement based
cryptography can be tomorrow's technology.

This work was supported by the Austrian Science Foundation FWF
(Projects No.~S6502, S6504 and F1506), the Austrian Academy of
Sciences, and the TMR program of the European Commission (Network
contract No.~ERBFMRXCT96-0087).

\begin{thebibliography}{99}
\bibitem{twopairs} Note also that in our case the beam splitter attack is less
effective than for coherent pulses, because even when two pairs
are produced simultaneously, Eve does not gain any information in
those cases where Alice and Bob detect photons belonging to the
same pair, because then the photon detected by Eve originates from
a different pair and is completely uncorrelated to Alice's and
Bob's photons.
\bibitem{bias} Note that it would be simple to bias the frequencies of analyzer
combinations to increase the production rate of the keys.
\bibitem{Bitmap}
Windows-BMP format containing $60 \times 90$ pixel, 8 bit color
information per pixel: $43200$ bit of picture information. The
file includes some header information and a color table, making
the entire picture file $51840$ bit. We encrypted only the
picture information, leaving the file header and the color table
unchanged.
\end{thebibliography}
\end{document}